\documentclass[graybox, vecphys]{svmult}
\usepackage{graphicx}        
\usepackage{multicol}        
\usepackage[bottom]{footmisc}\usepackage{newtxtext}      
\usepackage{newtxmath}       
\usepackage{url} 
\usepackage[latin1]{inputenc}

\usepackage{subeqnarray}
\usepackage{physics}
\usepackage{siunitx}
\usepackage{todonotes} 
\usepackage{comment}

\newcommand{\bs}{\boldsymbol}

\newcommand{\iu}{\mathrm{i}}
\DeclareMathOperator{\e}{e}

\makeindex             

\begin{document}

\title*{Multicomponent condensates: \\ a flexible platform for soliton physics}

\author{Franco Rabec, J\'er\^ome Beugnon, Jean Dalibard and Sylvain Nascimbene}

\institute{Laboratoire Kastler Brossel,  Coll\`ege de France, CNRS, ENS-PSL University, Sorbonne Universit\'e, 11 Place Marcelin Berthelot, 75005 Paris, France\\
\email{jean.dalibard@college-de-france.fr}}

\maketitle

\abstract
{\ In this chapter, we present a series of experimental investigations on binary mixtures of Bose-Einstein condensates. Our focus lies on the regime where the interaction parameters place the system at the threshold of miscibility. We demonstrate that the dynamics of such mixtures can be effectively reduced to a single nonlinear equation. This framework is illustrated through the discussion of stable solitonic solutions in one and two dimensions. Furthermore, we show that employing a binary mixture enables exploration beyond the dynamics governed by the nonlinear Schr\"odinger equation, allowing us to address other fundamental equations in nonlinear physics, such as the Landau-Lifshitz equation describing the motion of spin chains in ferromagnetic materials.}

\newpage


Mixtures of quantum fluids exhibit a remarkable variety of phenomena, depending on the statistical nature of their components and their miscibility properties. One of the earliest and still most significant examples is the mixture of liquid helium-3 and helium-4 \cite{leggett2006quantum}, which underlies the operation of the dilution refrigerator -- a unique cryogenic instrument that enables the attainment of millikelvin temperatures for macroscopic solid-state systems.

The achievement of gaseous Bose-Einstein condensation -- first in alkali-metal atoms and later in other atomic species such as alkaline-earth metals or lanthanides -- greatly expanded the range of mixtures accessible to experimental study \cite{baroni2024quantum,Varenna211}. These systems can involve two isotopes of the same element, such as lithium-6 and lithium-7 in their superfluid states \cite{ferrier2014mixture} forming a Bose--Fermi mixture analogous to the $^3$He-$^4$He  system mentioned above. Alternatively, mixtures may consist of two different atomic species, thereby allowing, for instance, the realization of heavy-light systems; a notable example is the Lithium-Ytterbium mixture \cite{hara2011quantum}, which exhibits a mass ratio close to 30 between its components.

In this Chapter, we focus on the opposite regime, where the constituents possess equal masses and correspond to different internal states of a single atomic species. The equality of masses  simplifies the theoretical description of the mixture and enables a meaningful parallel with other classes of quantum matter -- most notably, as will be discussed below, with the dynamics of ferromagnetic materials. We will restrict our attention to bosonic mixtures, leaving aside the equally important case of fermionic mixtures with opposite spins, which give rise to the well-known BEC-BCS crossover \cite{zwerger2011bcs}.

Employing various sublevels $\ket{1}, \ket{2}\ldots$ of the atomic ground state offers a wide range of possibilities for manipulating the mixture. For instance, the initial spatial distribution can be engineered by preparing all atoms in state $\ket{1}$ and subsequently transferring atoms to state $\ket{2}$ in a spatially resolved manner using appropriately shaped laser beams. Once this preparation is complete, the coupling between $\ket{1}$ and $\ket{2}$ can either be switched off -- this being the situation considered in what follows -- or kept active, thereby realizing coherently coupled condensates \cite{recati2022coherently}.

Mixtures involving multiple internal states of a single atomic species also provide a natural platform for exploring spinor physics \cite{Varenna211, stamper2013spinor}. For bosonic atoms with a ground electronic level characterized by a total spin $J \neq 0$, one can prepare samples in which all Zeeman sublevels $\ket{J,m}$ are populated. The resulting dynamics are remarkably rich, arising for example from $s$-wave elastic collisions in which a pair of atoms initially in states $(m_1, m_2)$ may scatter into states $(m_3, m_4)$, provided that the condition $m_1 + m_2 = m_3 + m_4$ is satisfied.

In this chapter, we focus on a two-component Bose-Einstein condensate (BEC), that is, a condensate composed of atoms from a single species occupying two distinct internal states, $\ket{1}$ and $\ket{2}$. For simplicity, we assume that no spin-exchange collisions occur, implying that once the system has been prepared, the individual populations of $\ket{1}$ and $\ket{2}$ remain conserved. Furthermore, we restrict our discussion to the mean-field regime, thereby excluding situations in which beyond-mean-field effects play a significant role (see \cite{petrov2015quantum,cabrera2018quantum}).

To keep the mathematical formulation as concise as possible, we restrict our analysis to the (experimentally relevant) one- and two-dimensional configurations. In these geometries, atoms are assumed to propagate freely either along the $x$ axis or within the $xy$ plane, while their motion in the orthogonal direction(s) is frozen by a strongly confining potential. Under such conditions, each component is described by a complex field $\psi_i$, whose dynamics are governed by two coupled nonlinear Schr\"odinger equations, usually called Gross-Pitaevskii equations (GPE) in this context: 
\begin{subeqnarray}
    \label{eq:2comp_gpe} \iu \hbar\pdv{\psi_1}{t} &=& -\frac{\hbar^2}{2M} \bs \nabla^2\psi_1
    +
    \left(g_{11} n_1 + g_{12}n_2\right)\psi_1 \\ \iu \hbar\pdv{\psi_2}{t} &=&
    -\frac{\hbar^2}{2M} \bs \nabla^2\psi_2
    + \left(g_{12}n_1+ g_{22} n_2\right) \psi_2\; ,
\end{subeqnarray} 
where $\hbar$ is the reduced Planck constant, $M$ is the atomic mass, $g_{11}$ and $g_{22}$ denote the intracomponent interaction parameters, $g_{12}$ the intercomponent coupling, and $n_i = \abs{\psi_i}^2$ the local densities of the two components. The interaction strengths $g_{ij}$ are proportional to the corresponding $s$-wave scattering lengths $a_{ij}$, and we assume throughout that all $g_{ij} > 0$, thereby avoiding collapse instabilities. We further define $N_i = \int |\psi_i(\bs r,t)|^2 \, \mathrm{d}^Dr$,  where $D$ is the system dimension, as the atom number in component $i$, and $n_t = n_1 + n_2$ as the total density.

The main objective of this chapter is to identify regimes in which the coupled GPE can be mapped onto well-known nonlinear equations involving a reduced number of fields, such as the single-component GPE or the Landau-Lifshitz equation (LLE). The physical implications of such reductions depend crucially on whether the mixture is miscible or immiscible. To characterize this property, we thus introduce the spin interaction parameter \cite{Timmermans1998}, 
\begin{equation} g_{s} = g - g_{12} \qquad \mbox{with}\qquad
g=\sqrt{g_{11}g_{22}}
\; . 
\label{eq:miscibility_criterion} 
\end{equation} 
The mixture is miscible for $g_s > 0$ and immiscible for $g_s < 0$. In the latter case, when both components initially occupy the entire available spatial region, a demixing instability develops \cite{Sadler06, Kronjager10, Tojo10, De14, Eto15, Jimenez19}. Rather than analyzing the instability mechanism itself, we focus here on the properties of the system once a steady state has been established.

A key assumption throughout this chapter is that the miscibility parameter $g_s$ is small compared to each individual coupling constant $g_{ij}$, that is, $|g_s| \ll g_{ij}$. This condition is well satisfied for alkali-metal species such as sodium and rubidium, and it carries important implications for the relevant energy scales of the mixture. Specifically, the pseudospin degree of freedom -- associated with the dynamics of the relative density $n_1 - n_2$ and governed by the coupling $g_s$ -- corresponds to an energy scale much smaller than that characterizing the dynamics of the total density $n_t$.

The structure of this Chapter is as follows. In Section~\ref{sec:def manakov}, we introduce the Manakov limit, where all interaction coefficients $g_{ij}$ can be taken as equal. In this case, we show that stable dark-bright solitons can be formed in 1D and we present their experimental observation. Section~\ref{sec:low_depletion} addresses the case of a nonzero and negative $g_s$ (i.e., the immiscible regime), under the assumption that the number of atoms in state $\ket{2}$ is sufficiently small for these atoms to form a localized droplet immersed in a weakly depleted background of atoms in state $\ket{1}$. We show that under these conditions, the problem can be reduced to a single GPE with an effective interaction strength related to $g_s$, and we illustrate this reduction through the formation of bright solitons in both one- and two-dimensional geometries. In Section~\ref{sec:arbitrary_depletion}, we lift the assumption of weak depletion of component 1 and discuss the cases $g_s > 0$ and $g_s < 0$ for a one-dimensional gas. We show that, at low energies, the system can be effectively described by a single-component nonlinear equation -- the Landau-Lifshitz equation -- which captures the dynamics of a spin chain with ferromagnetic interactions in an uniaxial medium. The two regimes, $g_s > 0$ and $g_s < 0$, correspond respectively to the easy-plane and easy-axis cases, and we present the realization of magnetic solitons in both of them. Finally, we summarize our findings in Section~\ref{sec:perspectives}.


\section{The Manakov regime}
\label{sec:def manakov}


\subsection{Definition of the regime}

The Manakov regime of the coupled GPE refers to the situation in which the two components have equal masses and nearly identical interaction parameters, \textit{i.e.}, $g_{11}\approx g_{22}$ and $|g_s|\ll g$. The Manakov limit corresponds to the case in which all the interaction parameters $g_{ij}$ are equal. While the Manakov regime may be achieved by carefully tuning the external magnetic field around Feshbach resonances, it arises naturally near zero magnetic field in certain alkali-metal species, such as rubidium and sodium atoms. 

In these atomic systems, the interaction potentials exhibit a weak singlet-triplet coupling. This results in scattering lengths close to each other for the magnetic sublevels of the two hyperfine ground manifolds, $F=1,2$. Consequently, binary mixtures prepared in these states naturally fall close to the Manakov regime. For experiments sensitive to magnetic perturbations, it is preferable to choose states with the same magnetic moment. Possible choices are either the pair of clock states $\ket{F=1, m_F=0}$, $\ket{F=2, m_F=0}$ \cite{Becker2008, Bakkali-Hassani2021}, or the pairs $\ket{F=1, m_F=\pm 1}$, $\ket{F=2, m_F=\mp 1}$ \cite{Tojo10, Nicklas11}. However, the lifetime of the $F=2$ level is limited by hyperfine relaxation, especially for sodium. It can then be more favorable to use mixtures within the $F=1$ manifold for experiments exploring long-time dynamics. For example, in the case of the mixture $\{\ket{1}=\ket{F=1,m_F=-1}$, $\ket{2}=\ket{F=1, m_F=1}\}$ for $^{23}\text{Na}$ atoms, the scattering lengths are $a_{11}=a_{22}\approx 54\, a_B$ and $a_{12}\approx 51\, a_B$~\cite{Knoop2011}, where $a_B$ is the Bohr radius. For the same mixture with $^{87}\text{Rb}$ atoms, the corresponding scattering lengths are $a_{11}=a_{22} \approx 100.4\, a_B$ and $a_{12} \approx 101.3\, a_B$~\cite{Vankempen2002}, very close to the Manakov limit.  Interestingly, the sign of $a_{12}-a_{11}$ differs for these two atoms, positioning the two mixtures on opposite sides of the miscibility transition: the Rb mixture is immiscible \cite{De14}, whereas the Na mixture is miscible \cite{Knoop2011}. In contrast, for other alkali-metal atoms (Li, K, $^{85}$Rb, Cs), the situation is more intricate, as Bose-Einstein condensation is typically achieved near a Feshbach resonance at a nonzero magnetic field.

In the Manakov regime, the low-energy dynamics can be effectively reduced to a pure spin dynamics. In more details, linearizing Eq.~\eqref{eq:2comp_gpe} excitations around a uniform equilibrium state yields a Bogoliubov excitation spectrum with two branches \cite{Timmermans98}. One branch is associated with a density mode, corresponding to sound waves at low momenta, and it is characterized by the absence of relative motion between the two components. The other branch is a spin mode, where the total density remains constant. For $g_s>0$, this branch also corresponds to sound waves at low momenta \cite{Kim20}, whereas it leads to the demixing instability for $g_s<0$. The typical energies associated  with density and spin modes involve respectively $g$ and $g_s$, implying a strong separation of the two branches. 


\subsection{The strict Manakov limit and dark-bright solitons}

The Manakov limit corresponds to the SU(2) symmetric case in which all interaction parameters $g_{ij}$ are equal, \textit{i.e.}, $g_s=0$. The coupled GPE introduced in Eq.~\eqref{eq:2comp_gpe}  then reduce to the Manakov system \cite{Manakov74}, originally formulated in the context of non-linear optics. In one dimension, this system is integrable and supports a rich variety of solitons including, for example, bright multi-solitons for attractive interactions \cite{Radhakrishnan95}.  

Among the various soliton families admitted by the Manakov model in 1D, a particularly remarkable solution is the dark-bright soliton introduced by Busch and Anglin \cite{Busch2001} in the context of ultracold quantum gases. This two-component soliton consists of a bright localized wave packet in one component, localized inside the density minimum of a dark soliton formed in the other component. At rest, the fields take the form $\psi_1(x) \propto \tanh (\kappa x)$ and $\psi_2(x)  \propto \cosh^{-1}(\kappa x)$, where the inverse width $\kappa$ is related to the total atom number $N_2$ in component~2 and the asymptotic density of component~1. As in the case of the dark soliton, the field $\psi_1(x)$ exhibits a characteristic $\pi$ phase jump at $x=0$ for a soliton at rest.

\begin{figure}
\includegraphics[width=\textwidth]{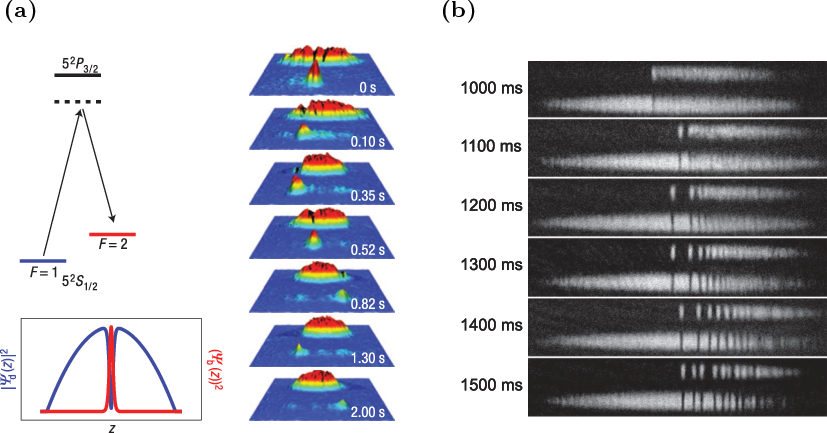}
\caption{\textbf{a} Realization of a single dark-bright soliton in a harmonic trap (adapted from Ref.~\cite{Becker2008}). A two-photon Raman transfer is used to spin-flip a wave packet of atoms from component $F=1$ to component $F=2$ in a spatially resolved way. A phase jump in component $F=1$, centered at the position of the wave packet, is also imprinted. The time evolution of  the dark-bright soliton in the trap is represented by a set of double-exposure absorption images showing the density distributions of both components over time. As expected in the Manakov regime, it can be seen that the total density remains approximately constant. \textbf{b} Realization of a soliton train of dark-bright solitons in a harmonic trap (adapted from Ref.~\cite{Hamner2011}). A magnetic field gradient induces a counterflow between the two components, resulting in the formation of solitons. For each time of evolution in the harmonic trap, the two pictures correspond to the two components of the gas.}
\label{Fig:Darkbright}
\end{figure}

The first experimental signatures of dark-bright solitons were reported in Ref.~\cite{Anderson01}. Here, we highlight two more recent experiments performed at Hamburg University and at Washington State University \cite{Becker2008,Hamner2011}. In both cases, the starting point of the experiments is a cigar-shaped trapped BEC of rubidium atoms. In the Hamburg experiment, a dark-bright soliton is imprinted optically using  spatially resolved two-photon Raman transfers from $\ket{1}\equiv\ket{F=1,m_F=0}$ to $\ket{2}\equiv\ket{F=2,m_F=0}$, which engineers both the $\pi$-phase jump in component~1 and the wave packet in component~2 with a controllable atom number (see Fig.\,\ref{Fig:Darkbright}a). A single dark-bright soliton is created and the authors observed  its oscillatory longitudinal motion in the harmonic trap at a frequency approximately 5 times smaller than the axial trapping frequency, in good agreement with the prediction of Ref.~\cite{Busch2001}. In the experiment at Washington State University, the authors start from a uniform mixture of  $\ket{F=1,m_F=1}$ and $\ket{F=2,m_F=2}$ atoms and trigger the formation of dark-bright solitons by inducing a counterflow instability. An external magnetic field gradient is applied to the mixture, generating opposite forces on the two components due to their opposite magnetic moment. The resulting relative motion becomes dynamically unstable and spontaneously leads to the formation of one or multiple dark-bright solitons, known as soliton trains (see Fig.\,\ref{Fig:Darkbright}b). 

\section{Low depletion limit and GPE solitons}
\label{sec:low_depletion}

In this section, we consider the case of an immiscible two-component mixture ($g_s<0$) in which the spin component~2 forms a localized wave packet immersed in a bath composed of spin component~1. We denote by $n_\infty$ the uniform bath density far from the minority component. When the number of minority atoms $N_2$ largely exceeds the characteristic value $N_2^{(0)} = n_\infty \xi_s^D$, where $\xi_s\equiv \hbar/\sqrt{2n_\infty M |g_s|}$ is the spin healing length, the minority species forms in the ground state a spherical spin domain with uniform density in its bulk. Here, we focus on the opposite limit $N_2 \ll N_2^{(0)}$, corresponding to a weak depletion regime $n_2 / n_1 \ll 1$. 


\subsection{Reduction to an attractive one-component model}
\label{section:attractive}

In the weakly-depleted regime, the two-component system can be reduced to a single-component model for the minority atoms, which interact through an effective interaction mediated by the bath \cite{Dutton05,Bakkali-Hassani2021,Bakkali23} (see also Ref.~\cite{Gliott25} for the derivation of an effective equation in the case of a balanced mixture)

Before addressing the macroscopic case, we start by considering two fixed minority atoms immersed in the bath \cite{Bakkali22}. Within the Fr\"ohlich Bose polaron framework, one can show that, in addition to their bare interaction, the two impurities experience a bath-mediated interaction arising from second-order coupling to the Bogoliubov excitations of the bath~\cite{Bakkali-Hassani2021}. The resulting effective interaction potential can be expressed in Fourier space as
\begin{equation}
V_{\text{eff}}(\mathbf{k}) = g_{22} - \frac{g_{12}^2}{g_{11}} \frac{1}{1 + \tfrac{1}{2}(k \xi)^2},
\end{equation}
where $\xi\equiv \hbar/\sqrt{2 M g n_\infty}$ is the healing length of the bath. In three dimensions, this corresponds to an attractive Yukawa potential, decaying over a characteristic range set by~$\xi$.

We now turn to the case of a macroscopic minority wave packet. In the Manakov regime,  the total density remains nearly constant,
\begin{equation}
n_1(\mathbf{r}) + n_2(\mathbf{r}) = n_\infty + \delta n(\mathbf{r}), \qquad |\delta n(\mathbf{r})| \ll n_\infty.
\end{equation}
At low energy, the spatial variations of each spin component occur over a typical length scale given by the spin healing length $\xi_s \gg \xi$ \cite{Timmermans1998}. Consequently, the mediated potential can be treated as short-ranged, and characterized by the effective coupling constant
\begin{equation}\label{eq:geff}
g_{\text{eff}} = g_{22} - \frac{g_{12}^2}{g_{11}}\approx 2\,g_s .
\end{equation}
In the immiscible regime, we have $g_{\text{eff}}<0$, i.e. the effective interaction is attractive.

The emergence of this attractive effective coupling can also be derived directly from the stationary regime of the coupled GPE (\ref{eq:2comp_gpe}) describing the spin mixture:
\begin{subeqnarray}
    \label{eq:2comp_gpe_static}
    \slabel{eq:2comp_gpe_static_1}
    \mu_1 \psi_1 &=& -\frac{\hbar^2}{2 M} \bs \nabla^2 \psi_1 + \left(g_{11} n_1  + g_{12} n_2\right) \psi_1, \\[2pt]
    \slabel{eq:2comp_gpe_static_2}
    \mu_2 \psi_2 &=& -\frac{\hbar^2}{2 M} \bs \nabla^2 \psi_2 + \left(g_{22} n_2  + g_{12} n_1\right) \psi_2,
\end{subeqnarray}
where $\mu_1$ and $\mu_2$ are the chemical potentials of the two components.  Expanding Eq.~\eqref{eq:2comp_gpe_static_1} in powers of $g_s / g_{ij}$, one finds at zeroth order the bath chemical potential $\mu_1 = g_{11} n_\infty$, consistent with a uniform density far from the minority atoms. At first order, the Laplacian term, of order $1/\xi_s^2$, can be approximated as $\nabla^2 \psi_1 \simeq \nabla^2 \sqrt{n_\infty - n_2}$, assuming a positive, real-valued wavefunction. Equation~\eqref{eq:2comp_gpe_static_1} then becomes
\begin{equation}
    g_{11} \, \delta n = \frac{\hbar^2}{2M}\frac{\bs \nabla^2 \sqrt{n_\infty - n_2}}{ \sqrt{n_\infty - n_2}}  + (g_{11} - g_{12}) n_2.
\end{equation}
Substituting this expression into Eq.~\eqref{eq:2comp_gpe_static_2} yields
\begin{equation}
    (\mu_2 - g_{12} n_\infty) \psi_2 = -\frac{\hbar^2}{2M} \bs \nabla^2 \psi_2 + g_{\text{eff}} |\psi_2|^2 \psi_2
    + \frac{\hbar^2}{2M} \frac{\bs \nabla^2 \sqrt{n_\infty - n_2}}{ \sqrt{n_\infty - n_2}}  \, \psi_2,
    \label{eq:eff_1comp_arbitrary_depletion}
\end{equation}
where $g_{\text{eff}} = g_{22} - g_{12}^2 / g_{11}$ is consistent with Eq.~\eqref{eq:geff}.  

So far, no assumption has been made about the depletion of component~1. Assuming now that the depletion is negligible, $n_2 / n_\infty \ll 1$, one obtains at lowest order
\begin{equation}
    \mu_{\text{eff}} \psi_2 = -\frac{\hbar^2}{2M} \bs \nabla^2 \psi_2 + g_{\text{eff}} |\psi_2|^2 \psi_2,
    \label{eq:eff_1comp_gpe}
\end{equation}
where $\mu_{\text{eff}} = \mu_2 - g_{12} n_\infty$ is the effective chemical potential.  Equation~\eqref{eq:eff_1comp_gpe} is precisely the GPE for a single-component Bose gas with attractive short-range interactions.

Compared to a genuine single-component attractive Bose gas, spin mixtures offer a key experimental advantage: they enable the preparation of arbitrary initial density profiles of the minority component through controlled spin transfer. Starting from a condensate fully polarized in state~$\ket{1}$, atoms can be locally transferred to state~$\ket{2}$ via a Raman transition driven by spatially modulated light fields \cite{Zou21}. This protocol benefits from the fact that the relevant length scale is the spin healing length $\xi_s$, typically $\xi_s \sim 5~\mu$m for mixtures in the Manakov regime, much larger than the usual healing length $\xi \sim 0.5~\mu$m.  

In the following subsections, we discuss several experimental realizations that exploit this spatial control over the minority component to produce different types of solitons.

\subsection{Realization of 1D bright solitons in a two-component gas}
\label{subsec:1D_bright_soliton}

In a 1D single-component, attractive Bose gas, the mean-field ground state of the system is a bright soliton with a density profile varying as $1/\cosh^2(\kappa x)$, where $\kappa$ is related to the interaction strength and to the atom number. The stability of this soliton results from  the competition between attractive interactions and single-particle dispersion. 

\begin{figure}[t]
    \sidecaption[t]
    \includegraphics[width=.5\textwidth]{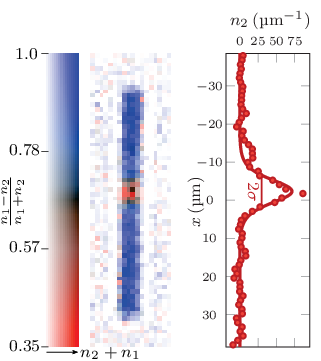}
    \caption{Spin-resolved density profile of a bright soliton realized in an immiscible spin mixture evolving in one dimension (adapted from~\cite{Rabec24}). The minority component forms a stationary localized wave packet within a bath of majority atoms.}
    \label{fig:bright_soliton}
\end{figure}

In this subsection, we transpose this concept to the case of a two-component quantum gas evolving in a one-dimensional geometry along $x$, under tight transverse confinement. In the weak-depletion limit, the density profile of the minority component takes the form
\begin{equation}
n_2(x) = \frac{N_2\kappa}{2} \frac{1}{\cosh^2(\kappa x)},
\end{equation}
where $N_2$ denotes the number of minority atoms, and the soliton characteristic width is given by $\kappa = M |g_\text{eff}| N_2/(2\hbar^2)$. Such magnetic solitons have been realized experimentally at Laboratoire Kastler Brossel in Paris using the immiscible spin mixture $\ket{1}=\ket{F=1,m_F=-1}$ and $\ket{2}=\ket{F=1, m_F=1}$ of $^{87}\mathrm{Rb}$ atoms~\cite{Rabec24}, as shown in Fig.~\ref{fig:bright_soliton}. The atoms are confined by a box-like optical potential with the shape of a tube, resulting in a uniform spatial density. The produced solitons constitute the two-component analog of the bright solitons observed in attractive single-component Bose gases~\cite{khaykovich2002formation,strecker2002formation}.

Beyond these stationary solitons, the focusing nonlinear Schr\"odinger equation also supports a wider class of remarkable solutions. Among them, the Peregrine soliton corresponds to a rational solution,
\begin{equation}
\psi_2(x,t) = \sqrt{n_{2\infty}} \left[ 1 - \frac{4(1 + 2 i t / \tau)}{1 + 4(x/\sigma)^2 + 4(t/\tau)^2} \right] e^{i t / \tau},
\end{equation}
which describes a wave packet localized in both space and time, emerging transiently from a uniform background of density $n_{2\infty}$. It thus represents a prototypical example of a rogue wave. The characteristic scales $\tau$ and $\sigma$ are related to the background density through
\begin{equation}
\frac{\hbar}{\tau} = \frac{\hbar^2}{M \sigma^2} = |g_\text{eff}| n_{2\infty}.
\end{equation}

\begin{figure}[t]
    \sidecaption[t]
    \includegraphics[width=7.5cm]{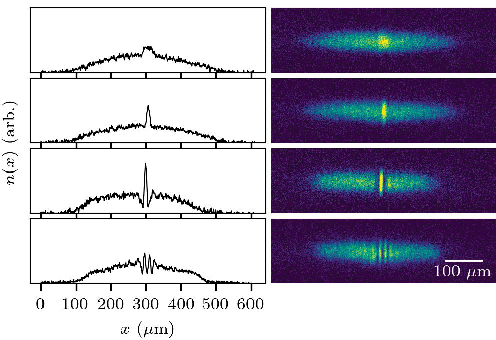}
    \caption{Dynamics of the minority component, starting with a quasi-uniform density background and following the application of a localized perturbation. The latter induces the formation of a Peregrine soliton, characterized by a transient density enhancement localized in both space and time (adapted from~\cite{Romero-Ros2024}).}
    \label{fig:peregrine}
\end{figure}

The Peregrine soliton has been experimentally realized using a spin mixture of $^{87}\mathrm{Rb}$ atoms~\cite{Romero-Ros2024}. In this experiment, a quasi-uniform immiscible spin mixture was prepared with a small ($\simeq15\%$) fraction in component~2. Upon applying a localized perturbation, the minority component exhibited a transient localized density enhancement consistent with the formation of a Peregrine soliton (see Fig.~\ref{fig:peregrine}).

\subsection{Realization of the 2D Townes soliton in a two-component gas}

We now turn to the case of spin mixtures evolving in two dimensions.  In this geometry, the nonlinear Schr\" odinger equation exhibits a remarkable scale invariance, which gives rise to a specific class of self-trapped stationary solutions known as the Townes solitons~\cite{chiao1964self}.  Townes solitons correspond to a family that can be expressed as
\begin{equation}\label{eq:Townes}
\psi_{N,\kappa}(\boldsymbol{r}) = \kappa \sqrt{N} \, \phi(\kappa r),
\end{equation}
where $\phi$ is a normalized and localized dimensionless function. The profile $\psi_{N,\kappa}(\boldsymbol{r})$ is stationary  for a critical atom number $N=N_{\text{c}} = A / |g|$, with $A \simeq 5.85$ and where $g<0$ is the interaction coupling constant. The fact that the stationary condition does not depend on  the characteristic width $\kappa^{-1}$ manifests the approximate scale invariance of the 2D Bose gas \cite{Pitaevskii97, Hung11,Yefsah11}. Conversely, a wave packet with an atom number $N>N_{\text{c}}$ undergoes collapse, while  it expands due to dispersion for $N<N_{\text{c}}$.  The Townes soliton therefore represents the marginally stable stationary solution separating these two regimes. It was observed in an attractive single-component cesium BEC at Purdue University~\cite{Chen2020}.  

\begin{figure}
\begin{center}
    \includegraphics[width=\textwidth]{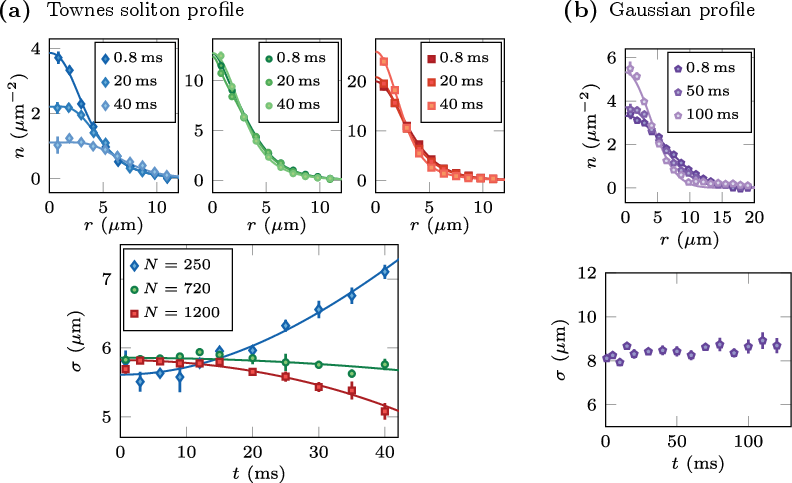}
    \caption{Dynamics of a localized wave packet of minority atoms evolving within a majority-atom bath in two dimensions. \textbf{a} When the initial density is prepared according to the Townes soliton profile, the wave packet remains stationary provided that the total atom number matches the critical value $N_{2,\text{c}} = A / |g_{\text{eff}}|$. \textbf{b} When the initial density instead follows a Gaussian profile, the root-mean-square width $\sigma$ can be made stationary \cite{Pitaevskii97}, but the density profile deforms over time (adapted from ~\cite{Bakkali-Hassani2021}).} 
    \label{fig:Townes}
\end{center}
\end{figure}

Here, we focus on the realization of a Townes soliton achieved by the Paris group with $^{87}\mathrm{Rb}$ atoms, using a hyperfine mixture of $\ket{1}=\ket{F=1,m_F=0}$ and $\ket{2}=\ket{F=2, m_F=0}$~\cite{Bakkali-Hassani2021}.  By preparing a localized minority wave packet with a total atom number close to the critical value $N_{2,\text{c}} = A / |g_{\text{eff}}|$, they observed a self-trapped state whose size remained constant over time  (see Fig.~\ref{fig:Townes}a). For $N_2 > N_{2, \text{c}}$, the expected contraction of the wave packet is observed. However, it does not result in a full collapse due to saturation effects at large depletion of the bath. Furthermore, they demonstrated that initializing the system with the Townes soliton profile given by Eq.~\eqref{eq:Townes} was crucial to obtain a fully stationary density distribution: other initial waveforms led to a deformation over time (see Fig.~\ref{fig:Townes}b).


\section{Arbitrary depletion case: Mapping to a ferromagnetic chain}
\label{sec:arbitrary_depletion} 

Ultracold quantum gases offer exceptionally versatile platforms for emulating diverse forms of quantum matter. In this section, we demonstrate that a binary mixture of Bose-Einstein condensates can serve as a simulator for the physics of ferromagnetic materials, as described by the Landau-Lifshitz equation. We focus on the one-dimensional case with $g_{11}=g_{22}$ and we illustrate this analogy through the formation of magnetic solitons, i.e., distinct localized structures that emerge from effective dipole-dipole interactions in a chain of magnetic moments.


\subsection{The Landau-Lifshitz equation}

We consider a one-dimensional chain of equally spaced magnetic moments, denoted by $\boldsymbol{\mu}_j$, with a nearest-neighbor interaction term $-J\,\boldsymbol{\mu}_j \cdot \boldsymbol{\mu}_{j+1}$, where $J>0$ corresponds to ferromagnetic coupling. In most materials, this isotropic interaction is complemented by a weaker anisotropic contribution, which we express as $-J'\,\mu_{j,z}\mu_{j+1,z}$ to describe an uniaxial material with rotational symmetry around the $z$-axis. The sign of $J'$ determines the preferred direction of magnetization: a positive value ($J'>0$) favors alignment along the $z$-axis (easy-axis anisotropy), whereas a negative value ($J'<0$) favors magnetization in the plane perpendicular to $z$ (easy-plane anisotropy).

By taking the continuum limit, $\boldsymbol{\mu}_j(t) \to \boldsymbol{M}(x,t)$, and choosing appropriate space and time units, we obtain that the vector field $\boldsymbol{M}(x,t)$ satisfies the Landau-Lifshitz equation \cite{Landau1935, Landau1980}:
\begin{equation}
\pdv{\boldsymbol{M}}{t}
= \left(\pdv[2]{\boldsymbol{M}}{x} \pm M_z\,\boldsymbol{e_z}\right) \times \boldsymbol{M}\ ,
\label{eq:lle}
\end{equation}
where the $+$ sign corresponds to the easy-axis case and the $-$ sign to the easy-plane case. Without loss of generality, we can assume that the vector field $\boldsymbol{M}(x,t)$ has unit norm at every point, i.e., $|\boldsymbol{M}(x,t)| = 1$. This equation plays a central role in the theory of magnetic materials.  It describes the form of the interface between magnetic domains as well as more complex static and dynamical structures within these systems.

The connection with the binary mixture problem can be established by starting from the coupled equations~(\ref{eq:2comp_gpe}) and simplifying them in the Manakov limit, under the assumption that the total density $n_t=n_1 + n_2$ remains uniform and time-independent. Denoting this constant density by $n_0$, the spinor $(\psi_1, \psi_2)$ can then be parametrized in terms of three fields: the mixing angle $\theta$, the global phase $\Phi/2$, and the relative phase $\varphi$:
\begin{equation}
\begin{pmatrix}
	\psi_1 \\
	\psi_2
\end{pmatrix} = \sqrt{n_0} \e^{\iu \Phi/2} \begin{pmatrix}
	\cos(\theta/2)\e^{-\iu \varphi/2}\\
	\sin(\theta/2)\e^{+\iu \varphi/2}
    \label{eq:param_spherical}
\end{pmatrix}.
\end{equation}
We can first write Eqs.\,(\ref{eq:2comp_gpe}) in terms of these three fields and then take advantage of the equation of continuity, which allows us to link them by $\partial_{x}\Phi - \cos(\theta)\,\partial_{x} \varphi = I$, where $I$ is the (dimensionless) current. Since $I$ is independent of space and time, we can use this relation to eliminate $\Phi$ from the equations of evolution. In the following, we make the additional assumption $I=0$, since we will be focusing on systems with strict boundary conditions (for solutions with $I\neq 0$, see e.g. \cite{Rabec24} and refs. in). 

Finally, we introduce the vector field 
\begin{equation}
\bs M(x,t)=-(\cos\varphi\sin\theta,\sin\varphi\sin\theta,\cos\theta)^T
\end{equation}
and verify that it satisfies the Landau-Lifshitz equation, using dimensionless time and length units given in Table~\ref{tab:magnetic_soliton_dimensionless}. This correspondence thereby establishes the connection between the two-fluid system and the magnetic chain. More precisely, the easy-axis configuration [the $+$ sign in Eq.~\eqref{eq:lle}] corresponds to the immiscible regime, while the easy-plane configuration corresponds to the miscible regime. This mapping extends the reduction to an effective one-component model for an immiscible mixture of \S\,\ref{section:attractive} (see Eq.~\eqref{eq:eff_1comp_arbitrary_depletion}), by removing the restriction that the wavefunction $\psi_1$ must be real and positive.

\begin{table}[!t]
\caption{Scaling factors used in this Chapter to switch from dimensionless to physical units.}
\label{tab:magnetic_soliton_dimensionless}
\centering
\renewcommand{\arraystretch}{1.5}
\begin{tabular}{p{3cm}l@{}}
\hline\noalign{\smallskip}
Physical quantity & Scaling factor \\
\noalign{\smallskip}\svhline\noalign{\smallskip}
Distance & $\xi_s = \hbar/\sqrt{2n_0 M \abs{g_s}}$  \\
Time & $\tau_s = (2 M \xi_s^2)/\hbar $ \\
Velocity & $c_s=\xi_s/\tau_s$ \\
Momentum &  $\hbar n_0$ \\
Energy & $\hbar n_0 c_s$ \\
\noalign{\smallskip}\hline\noalign{\smallskip}
\end{tabular}
\end{table}


\subsection{A particular solution: The magnetic soliton}

The Landau-Lifshitz equation, like most nonlinear equations in physics, admits a wide variety of solitonic and breathing solutions. Here, we focus on the simplest localized structure, the fundamental magnetic soliton, which can arise in both the miscible and immiscible regimes, but with a notably different spatial arrangement. We assume that the two fluids are confined within a segment of length $L$ much larger than the soliton size, allowing us to neglect edge effects. For convenience, we consider a solution centered in $x = 0$ at $t = 0$; owing to the translational invariance of Eqs.~(\ref{eq:2comp_gpe}) and~(\ref{eq:lle}), this solution can be straightforwardly generalized to an arbitrary position at any given time.


\vskip 2mm\noindent \textbf{The easy-plane case.} This configuration has been studied in detail in Refs.~\cite{Qu2016,Congy2016}. We briefly summarize here the main results for the case of asymptotically equal densities, $n_1 = n_2 = n_0/2$. Assuming, for instance, a slight excess of atoms in component~1 relative to component~2, the mixing angle $\theta$ and the relative phase $\varphi$ for a soliton moving with velocity~$v$ are given at $t = 0$ by
\begin{equation}
\cos [\theta(x)]=\frac{\sqrt{1-v^2}}{\cosh\left(x\sqrt{1-v^2}\right)}\qquad \qquad \tan [\varphi(x)]=\frac{v}{\sinh\left(x\sqrt{1-v^2}\right)}\ .
\end{equation} 
The mixing angle $\theta$ tends to $\pi/2$ as $x \to \pm \infty$; it remains positive everywhere and reaches its minimum at $x = 0$. This corresponds to a density bump in component~1, $n_1(x)$, around the soliton center, accompanied by a corresponding dip in $n_2(x)$, such that the total density $n_1(x) + n_2(x)$ remains uniform and equal to $n_0$. The characteristic width of this structure scales as $(1 - v^{2})^{-1/2}$, with the physical units of position and velocity given in Table~\ref{tab:magnetic_soliton_dimensionless}. Full depletion of component~2 ($n_2(0) = 0$, hence $\cos[\theta(0)] = 1$) occurs only for a soliton at rest.

The relative phase $\varphi$ varies continuously from $0$ as $x \to -\infty$ to $\pi$ as $x \to +\infty$, except in the stationary case ($v = 0$), where $\varphi(x)$ exhibits a $\pi$ phase jump at the origin. In this particular case, the wavefunctions $\psi_1$ and $\psi_2$ can both be taken real, with $\psi_1(x)$ positive everywhere and $\psi_2(x)$ possessing a node at $x = 0$. Note that in the miscible case, the ground state of the system corresponds to a uniform distribution of atoms in both components, so this solitonic configuration represents an excited state of the binary mixture, with energy $\sqrt{1 - v^{2}}$.


\vskip 2mm\noindent \textbf{The easy-axis case.} In the immiscible regime, we consider a finite number $N_2$ of atoms of species~2 immersed in a (possibly locally fully-depleted) bath of species~1. In this case, the soliton can represent the ground state of the binary mixture: atoms of species~2 form a droplet whose size results from the competition between interaction and kinetic energies. For simplicity, we focus here on the stationary case of a soliton at rest. The corresponding solutions are given by
\begin{equation}
\frac{n_2(x)}{n_0}=\sin^2\frac{\theta(x)}{2}=\frac{2\kappa^2}{1+\kappa^2 +|1-\kappa^2|\cosh(2\kappa x)},
\end{equation} 
where the real and positive parameter $\kappa$ is related to $N_2$ by
\begin{equation}
\frac{N_2}{n_0 \xi_s}=\ln\frac{1+\kappa}{|1-\kappa|}.
\label{eq:link_N2_kappa}
\end{equation} 
For a given $N_2$, Eq.~\eqref{eq:link_N2_kappa} admits two possible values of~$\kappa$. The first  one, $\kappa < 1$, corresponds to the ground state of the system. In this case, the two phases $\Phi$ and $\varphi$ are spatially uniform, so that the binary mixture can be described by two real, nodeless wavefunctions $\psi_1(x)$ and $\psi_2(x)$. The second solution, $\kappa > 1$, corresponds to a narrower soliton and it is an excited state of the system. Here, $\psi_2(x)$ can be chosen real and positive everywhere, while $\psi_1(x)$ exhibits a node at $x = 0$, reminiscent of the dark-bright solitons discussed in Sec.~\ref{sec:def manakov}. This analogy is, however, only qualitative, since the characteristic length scale $\xi_s$, which sets the magnetic soliton size, does not appear in the dark-bright soliton configuration.

For the general case of a moving magnetic soliton in the  immiscible regime, we refer the reader to Ref.~\cite{Rabec24} and references therein. Here, we simply recall the soliton's dispersion relation, which connects its energy~$E$ to its canonical momentum~$P$:
\begin{equation}
E(P)= E(0) + 4\beta \sin^2(P/2)\qquad \mbox{with}\quad \beta^{-1}=\sinh(N_2/n_0\xi_s)\ ,
\label{eq:energy_momentum_easy_axis}
\end{equation}
from which we deduce the relation between the soliton velocity and its momentum:
\begin{equation}
v=\frac{\partial E}{\partial P}= 2\beta\sin P\ .
\end{equation}
The periodic dispersion relation~(\ref{eq:energy_momentum_easy_axis}) -- reminiscent of that of a quantum particle moving in an external periodic potential -- underlies the remarkable properties of magnetic solitons in the easy-axis regime (see Refs.~\cite{Rabec24, Kosevich98, Bresolin23} and Sec.~\ref{sec:static_magnetic_soliton}).

 
\subsection{Experimental studies of the easy-plane solitonic solution}

\begin{figure}
    \sidecaption
    \includegraphics[width=0.5\textwidth]{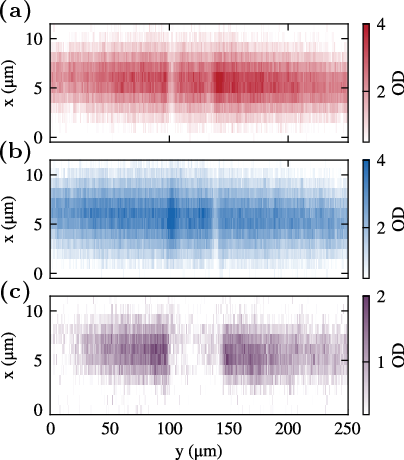} 
    \caption{Measurement of the density and phase profiles of a pair of easy-plane magnetic solitons. The $y$ coordinate corresponds to the longitudinal direction (i.e., the $x$ coordinate in the main text). \textbf{a} Optical density (OD) of component~1. \textbf{b} Optical density of component~2. \textbf{c} Measurement of the relative phase $\varphi(x)$. A Ramsey interferometric scheme maps $\cos{ \varphi(x) }$ to the optical density.  The relative phase changes by $\pi$ across each soliton, with opposite phase evolution for the two solitons (adapted from \cite{Farolfi2020}).}
\label{fig:easy_plane_exp_profile}
\end{figure}

Easy-plane magnetic solitons were realized experimentally in a $^{23}\text{Na}$ BEC in Trento University \cite{Farolfi2020} and at Georgia Tech \cite{Chai2020}. These experiments were performed with a quasi-1D gas confined in an axial harmonic potential, resulting in a cigar-shaped cloud. The solitons were produced using the miscible spin mixture $\ket{1}=\ket{F=1,m_F=-1}$ and $\ket{2}=\ket{F=1, m_F=1}$. Starting from a balanced mixture of both components, a laser beam generates a step-like potential that imprints a relative phase difference between the two halves of the cloud. The laser beam is engineered to induce opposite phase differences for the two components. It is applied during a sufficiently short duration so that it does not affect the spatial density. As a result, the total magnetization $\cos\theta\propto n_2 - n_1$ remains zero everywhere. Because magnetization is a conserved quantity, this protocol generates pairs of magnetic solitons of opposite magnetization. The finite resolution of the imaging system smoothens the step-like potential, resulting in a finite width of the phase jump and thus a finite soliton velocity. The two solitons created in this way propagate in opposite directions due to their opposite magnetization. The measured density is shown in Fig.~\ref{fig:easy_plane_exp_profile}.

\begin{figure}[t]
    \sidecaption[t]
    \includegraphics[width=0.5\textwidth]{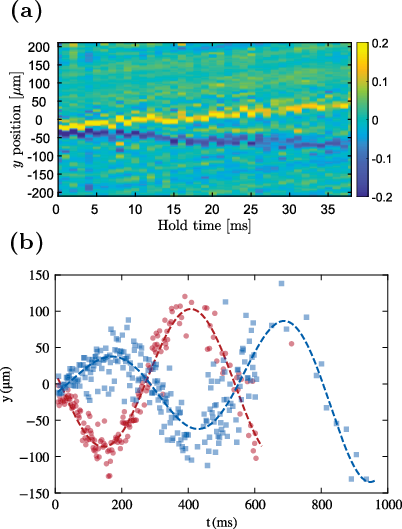}
    \caption{\label{fig:easy_plane_exp_dynamics}Evolution of the positions of the two solitons in the harmonic trap, the $y$ coordinate corresponds to the longitudinal direction (i.e. the $x$ coordinate in the main text). \textbf{a} Magnetization profile versus time at short times, when the solitons propagate at constant velocity. \textbf{b} Position of the two solitons at longer times, showing oscillatory motion in the trap (adapted respectively from \cite{Chai2020} and \cite{Farolfi2020}).} 
\end{figure}

At short times ($<\,\SI{50}{\milli\second}$), when the curvature of the trap can be neglected, the two solitons propagate at approximately constant velocity. Additionally, the relative phase $\varphi$ is measured through Ramsey interferometry, and a phase difference of $\pi$ across the soliton is observed (see Fig.~\ref{fig:easy_plane_exp_profile}). At longer times, the curvature of the trap comes into play, and oscillations of the solitons in the harmonic trap are observed \cite{Farolfi2020}. The dynamics of a soliton in the trap can be derived from energy conservation within the local-density approximation \cite{Qu2016}. The measured dynamics at short and long times are shown in Fig.~\ref{fig:easy_plane_exp_dynamics}.

By changing the imprinted potential from a step function to a top-hat function, two pairs of solitons are created \cite{Farolfi2020, Chai2020}, and collisions between solitons having the same or opposite magnetization sign are observed.

\subsection{Experimental studies of the easy-axis solitonic solution}
\label{sec:static_magnetic_soliton}

Easy-axis magnetic solitons were realized experimentally by the Paris group with a $^{87}\text{Rb}$ BEC \cite{Rabec24} using the same setup as in \S\,\ref{subsec:1D_bright_soliton}. Starting from a cloud initially prepared in state $\ket{1}$, a spatially resolved spin transfer  provides the desired magnetization profile while maintaining a uniform phase $\varphi$ and a uniform total density. The profile of a magnetic soliton at rest with arbitrary depletion, corresponding to the solution with $\kappa < 1$, can then be directly imprinted on the atoms. 

\begin{figure}
    \includegraphics[width=\textwidth]{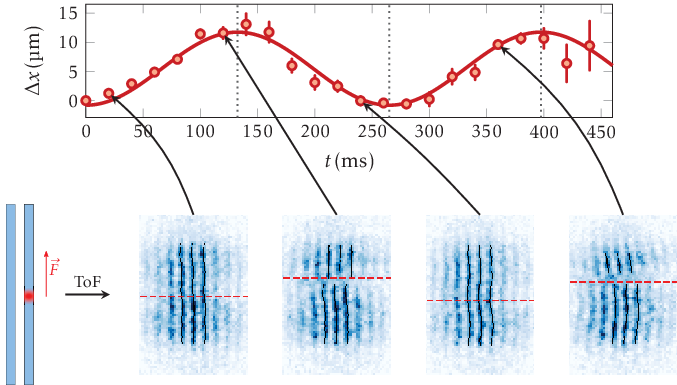}    
    \caption{Evolution of the position of the soliton submitted to a constant force, showing oscillatory motion reminiscent of Bloch oscillations. A second BEC, containing only atoms in component~1 provides a phase reference for an interferometric measurement of the phase of the bath. The interferograms, displayed below, show no phase jump across the soliton at $t\equiv 0 \pmod T$, while a $\pi$ phase jump is observed at $t\equiv T/2 \pmod T$. Adapted from \cite{Rabec24, Rabec25a}.}
    \label{fig:Bloch oscillations}
\end{figure}

Then, by applying a force $F$ that acts differently on components 1 and 2, a canonical momentum $P$ can be imparted to the soliton according to Newton's equation $\dot P=F$. If the force is small enough, the evolution is adiabatic and explores the full family of solitonic states at constant $N_2$. Since the soliton energy is a periodic function of its momentum [see Eq.~\eqref{eq:energy_momentum_easy_axis}], the evolution is oscillatory, as observed experimentally in \cite{Rabec24} (see Fig.~\ref{fig:Bloch oscillations}). In particular, after half a period of oscillation, the soliton is again at rest, but with $\kappa>1$. The phase of component~1 is measured by interfering it with a reference BEC. No phase jump is observed across the soliton initially, confirming that the initial state of the system corresponds to the solution with $\kappa<1$. After half a period of oscillation, a $\pi$ phase jump is observed, a signature of the $\kappa>1$ solution.

\section{Conclusions}
\label{sec:perspectives}

In this Chapter, we have shown that multicomponent Bose-Einstein condensates provide highly versatile platforms for addressing various emblematic problems in nonlinear physics. We focused on binary mixtures, whether miscible or immiscible, and explored different solitonic structures that may arise. Our analysis mainly concentrated on the regime where the spin interaction strength $g_s$ is small compared to interspecies and intraspecies interaction strengths.

In the limit where one of the two species is only weakly depleted by the presence of the other, we showed that the problem can be described by an effective single-component Gross-Pitaevskii equation. When the depletion of each component can become arbitrarily large, the system can instead be mapped onto another well-known equation of nonlinear physics, the Landau-Lifshitz equation, originally introduced to study the dynamics of magnetization in ferromagnetic media.

This list of multicomponent problems reducible to an effective single-component description is not exhaustive. Another remarkable example concerns the sine-Gordon equation, another fundamental equation in nonlinear physics. This equation can, for instance, be recovered from the Landau-Lifshitz equation in the easy-axis case by applying an external magnetic field in a direction orthogonal to the material's axis \cite{dauxois2006physics}, a situation that can be mapped on the binary mixture considered here by applying a coherent coupling between the two spin states. An experimental simulator of this equation was realized by \cite{schweigler2017experimental} using a pair of one-dimensional condensates in the presence of tunnel-coupling between them.

A major advantage of using mixtures of Bose-Einstein condensate to realize these different hallmark situations lies in the flexibility of such systems. As we have shown, using light beams to coherently transfer atoms from one component to another enables a fine control over the preparation of the desired state, whether it be a fundamental soliton or a more complex structure such as the Peregrine soliton. Moreover, one is not limited to the preparation of a single object: multiple structures can be generated simultaneously to study their interactions and collision dynamics. The use of Feshbach resonances further allows real-time tuning of the interaction strength and even its nature, whether attractive or repulsive. These condensate mixtures therefore constitute an extremely flexible platform, opening connections to many other areas of nonlinear physics, and we hope that this Chapter will serve as an invitation to pursue this exploration further.

\section*{Acknowledgements}
This preprint will appear as a chapter in the Springer book entitled \textit{Short and Long Range Quantum Atomic Platforms --- Theoretical and Experimental Developments} (provisional title), edited by P.~G.~Kevrekidis, C.~L.~Hung, and S.~I.~Mistakidis.

\bibliographystyle{spphys}

\begin{thebibliography}{10}
\providecommand{\url}[1]{{#1}}
\providecommand{\urlprefix}{URL }
\expandafter\ifx\csname urlstyle\endcsname\relax
  \providecommand{\doi}[1]{DOI \discretionary{}{}{}#1}\else
  \providecommand{\doi}{DOI \discretionary{}{}{}\begingroup \urlstyle{rm}\Url}\fi

\bibitem{leggett2006quantum}
A.J. Leggett, \emph{Quantum liquids: Bose condensation and Cooper pairing in condensed-matter systems} (Oxford university press, 2006)

\bibitem{baroni2024quantum}
C.~Baroni, G.~Lamporesi, M.~Zaccanti, Nat. Rev. Phys. \textbf{6}(12), 736 (2024)

\bibitem{Varenna211}
R.~Grimm, M.~Inguscio, S.~Stringari, G.~Lamporesi (eds.),
\newblock \emph{Quantum Mixtures with Ultra-cold Atoms}, \emph{Proceedings of the International School of Physics ``Enrico Fermi''}, vol. 211 (IOS Press, Amsterdam, 2025)

\bibitem{ferrier2014mixture}
I.~Ferrier-Barbut, M.~Delehaye, S.~Laurent, A.T. Grier, M.~Pierce, B.S. Rem, F.~Chevy, C.~Salomon, Science \textbf{345}(6200), 1035 (2014)

\bibitem{hara2011quantum}
H.~Hara, Y.~Takasu, Y.~Yamaoka, J.M. Doyle, Y.~Takahashi, Phys. Rev. Lett. \textbf{106}(20), 205304 (2011)

\bibitem{zwerger2011bcs}
W.~Zwerger, \emph{The BCS-BEC crossover and the unitary Fermi gas}, vol. 836 (Springer Science \& Business Media, 2011)

\bibitem{recati2022coherently}
A.~Recati, S.~Stringari, Annu. Rev. Condens. Matter Phys. \textbf{13}(1), 407 (2022)

\bibitem{stamper2013spinor}
D.M. Stamper-Kurn, M.~Ueda, Rev. Mod. Phys. \textbf{85}(3), 1191 (2013)

\bibitem{petrov2015quantum}
D.S. Petrov, Phys. Rev. Lett. \textbf{115}(15), 155302 (2015)

\bibitem{cabrera2018quantum}
C.R. Cabrera, L.~Tanzi, J.~Sanz, B.~Naylor, P.~Thomas, P.~Cheiney, L.~Tarruell, Science \textbf{359}(6373), 301 (2018)

\bibitem{Timmermans1998}
E.~Timmermans, Phys. Rev. Lett. \textbf{81}(26), 5718 (1998)

\bibitem{Sadler06}
L.E. Sadler, J.M. Higbie, S.R. Leslie, M.~Vengalattore, D.M. Stamper-Kurn, Nature \textbf{443}(7109), 312 (2006)

\bibitem{Kronjager10}
J.~Kronj\"ager, C.~Becker, P.~Soltan-Panahi, K.~Bongs, K.~Sengstock, Phys. Rev. Lett. \textbf{105}(9), 090402 (2010)

\bibitem{Tojo10}
S.~Tojo, Y.~Taguchi, Y.~Masuyama, T.~Hayashi, H.~Saito, T.~Hirano, Phys. Rev. A \textbf{82}(3), 033609 (2010)

\bibitem{De14}
S.~De, D.L. Campbell, R.M. Price, A.~Putra, B.M. Anderson, I.B. Spielman, Phys. Rev. A \textbf{89}(3), 033631 (2014)

\bibitem{Eto15}
Y.~Eto, M.~Kunimi, H.~Tokita, H.~Saito, T.~Hirano, Phys. Rev. A \textbf{92}(1), 013611 (2015)

\bibitem{Jimenez19}
K.~Jim{\'e}nez-Garc{\'\i}a, A.~Invernizzi, B.~Evrard, C.~Frapolli, J.~Dalibard, F.~Gerbier, Nat. Commun. \textbf{10}(1), 1422 (2019)

\bibitem{Becker2008}
C.~Becker, S.~Stellmer, P.~{Soltan-Panahi}, S.~D{\"o}rscher, M.~Baumert, E.M. Richter, J.~Kronj{\"a}ger, K.~Bongs, K.~Sengstock, Nat. Phys. \textbf{4}(6), 496 (2008)

\bibitem{Bakkali-Hassani2021}
B.~{Bakkali-Hassani}, C.~Maury, {Y.-Q. Zou}, {\'E}.~Le~Cerf, R.~{Saint-Jalm}, P.C.M. Castilho, S.~Nascimbene, J.~Dalibard, J.~Beugnon, Phys. Rev. Lett. \textbf{127}(2), 023603 (2021)

\bibitem{Nicklas11}
E.~Nicklas, H.~Strobel, T.~Zibold, C.~Gross, B.A. Malomed, P.G. Kevrekidis, M.K. Oberthaler, Phys. Rev. Lett. \textbf{107}(19), 193001

\bibitem{Knoop2011}
S.~Knoop, T.~Schuster, R.~Scelle, A.~Trautmann, J.~Appmeier, M.K. Oberthaler, E.~Tiesinga, E.~Tiemann, Phys. Rev. A \textbf{83}(4), 042704 (2011)

\bibitem{Vankempen2002}
E.G.M. {van Kempen}, S.J.J.M.F. Kokkelmans, D.J. Heinzen, B.J. Verhaar, Phys. Rev. Lett. \textbf{88}(9), 093201 (2002)

\bibitem{Timmermans98}
E.~Timmermans, Phys. Rev. Lett. \textbf{81}(26), 5718 (1998)

\bibitem{Kim20}
J.H. Kim, D.~Hong, Y.~Shin, Phys. Rev. A \textbf{101}(6), 061601 (2020)

\bibitem{Manakov74}
S.V. Manakov, Sov. Phys. JETP \textbf{38}(2), 248 (1974)

\bibitem{Radhakrishnan95}
R.~Radhakrishnan, M.~Lakshmanan, J. Phys. A: Math. Gen. \textbf{28}(9), 2683 (1995)

\bibitem{Busch2001}
{\relax Th}.~Busch, J.R. Anglin, Phys. Rev. Lett. \textbf{87}(1), 010401 (2001)

\bibitem{Anderson01}
B.P. Anderson, P.C. Haljan, C.A. Regal, D.L. Feder, L.A. Collins, C.W. Clark, E.A. Cornell, Phys. Rev. Lett. \textbf{86}(14), 2926 (2001)

\bibitem{Hamner2011}
C.~Hamner, J.J. Chang, P.~Engels, M.A. Hoefer, Phys. Rev. Lett. \textbf{106}(6), 065302 (2011)

\bibitem{Dutton05}
Z.~Dutton, C.W. Clark, Phys. Rev. A \textbf{71}(6), 063618 (2005)

\bibitem{Bakkali23}
B.~Bakkali-Hassani, C.~Maury, S.~Stringari, S.~Nascimbene, J.~Dalibard, J.~Beugnon, New J. Phys. \textbf{25}, 013007 (2023)

\bibitem{Gliott25}
E.~Gliott, C.~Piekarski, N.~Cherroret, Phys. Rev. Res. \textbf{7}(3), 033189 (2025)

\bibitem{Bakkali22}
B.~Bakkali-Hassani, J.~Dalibard, in \emph{Quantum {{Mixtures}} with {{Ultra-cold Atoms}}}, Proceedings of the International School of Physics "Enrico Fermi", Vol.~211 (IOS Press, Amsterdam, 2025), chap. Townes Soliton and beyond: {{Non-miscible Bose}} Mixtures in {{2D}}

\bibitem{Zou21}
{Y.-Q. Zou}, {\'E}.~Le~Cerf, B.~Bakkali-Hassani, C.~Maury, G.~Chauveau, P.C.M. Castilho, R.~Saint-Jalm, S.~Nascimbene, J.~Dalibard, J.~Beugnon, J.Phys. B : At. Mol. Opt. \textbf{54}(8), 08LT01 (2021)

\bibitem{Rabec24}
F.~Rabec, G.~Chauveau, G.~Brochier, S.~Nascimbene, J.~Dalibard, J.~Beugnon, Nat. Phys. \textbf{21}(10), 1541 (2025)

\bibitem{khaykovich2002formation}
L.~Khaykovich, F.~Schreck, G.~Ferrari, T.~Bourdel, J.~Cubizolles, L.D. Carr, Y.~Castin, C.~Salomon, Science \textbf{296}(5571), 1290 (2002)

\bibitem{strecker2002formation}
K.E. Strecker, G.B. Partridge, A.G. Truscott, R.G. Hulet, Nature \textbf{417}(6885), 150 (2002)

\bibitem{Romero-Ros2024}
A.~{Romero-Ros}, G.C. Katsimiga, S.I. Mistakidis, S.~Mossman, G.~Biondini, P.~Schmelcher, P.~Engels, P.G. Kevrekidis, Phys. Rev. Lett. \textbf{132}(3), 033402 (2024)

\bibitem{chiao1964self}
R.Y. Chiao, E.~Garmire, C.H. Townes, Phys. Rev. Lett. \textbf{13}(15), 479 (1964)

\bibitem{Pitaevskii97}
L.P. Pitaevskii, A.~Rosch, Phys. Rev. A \textbf{55}(2), R853 (1997)

\bibitem{Hung11}
C.L. Hung, X.~Zhang, N.~Gemelke, C.~Chin, Nature \textbf{470}(7333), 236 (2011)

\bibitem{Yefsah11}
T.~Yefsah, R.~Desbuquois, L.~Chomaz, K.J. G{\"u}nter, J.~Dalibard, Phys. Rev. Lett. \textbf{107}(13), 130401 (2011)

\bibitem{Chen2020}
C.A. Chen, C.L. Hung, Phys. Rev. Lett. \textbf{125}(25), 250401 (2020)

\bibitem{Landau1935}
L.D. Landau, E.~Lifshitz, Phys. Z. Sowjetunion \textbf{8}(153), 101 (1935)

\bibitem{Landau1980}
L.D. Landau, E.M. Lifshitz, L.P. Pitaevskii, \emph{Statistical {{Physics}}, {{Part}} 2}, \emph{Course of {{Theoretical Physics}}}, vol.~9 (Pergamon Press, 1980)

\bibitem{Qu2016}
C.~Qu, L.P. Pitaevskii, S.~Stringari, Phys. Rev. Lett. \textbf{116}(16), 160402 (2016)

\bibitem{Congy2016}
T.~Congy, A.~Kamchatnov, N.~Pavloff, SciPost Phys. \textbf{1}(1), 006 (2016)

\bibitem{Kosevich98}
A.M. Kosevich, V.V. Gann, A.I. Zhukov, V.P. Voronov, J. Exp. Theor. Phys. \textbf{87}, 401 (1998)

\bibitem{Bresolin23}
S.~Bresolin, A.~Roy, G.~Ferrari, A.~Recati, N.~Pavloff, Physical Review Letters \textbf{130}(22), 220403 (2023)

\bibitem{Farolfi2020}
A.~Farolfi, D.~Trypogeorgos, C.~Mordini, G.~Lamporesi, G.~Ferrari, Phys. Rev. Lett. \textbf{125}(3), 030401 (2020)

\bibitem{Chai2020}
X.~Chai, D.~Lao, K.~Fujimoto, R.~Hamazaki, M.~Ueda, C.~Raman, Phys. Rev. Lett. \textbf{125}(3), 030402 (2020)

\bibitem{Rabec25a}
F.~Rabec, Inducing {P}eriodic {E}ffects in {B}ose-{E}instein {C}ondensates: {S}oliton {O}scillations and {S}uperfluid {F}raction.
\newblock Ph.D. thesis, Sorbonne Universit{\'e} (2025)

\bibitem{dauxois2006physics}
T.~Dauxois, M.~Peyrard, \emph{Physics of solitons} (Cambridge University Press, 2006)

\bibitem{schweigler2017experimental}
T.~Schweigler, V.~Kasper, S.~Erne, I.~Mazets, B.~Rauer, F.~Cataldini, T.~Langen, T.~Gasenzer, J.~Berges, J.~Schmiedmayer, Nature \textbf{545}(7654), 323 (2017)

\end{thebibliography}

\end{document}